# Random Lasing from Dyed Polystyrene Spheres in Disordered Environment

*Sunita Kedia and Sucharita Sinha*

*Laser and Plasma Surface Processing Section, Bhabha Atomic Research Centre, Mumbai 400 085, India*

*Email: skedia@barc.gov.in

**Abstract:** Advance designs of random lasers towards development of miniature laser systems are in demand. We demonstrated random lasing from Rhodamine-B dye attached to polystyrene micro-spheres. Bare polystyrene spheres were used as scatterers and these provided optical feedback to the gain. Random lasing was successfully demonstrated in two different disordered environments, in binary colloidal mixture solution and in photonic glass. Incoherent feedback occurred in both the cases and single wavelength lasing were obtained. The lasing threshold in case of photonic glass was lower in comparison to binary colloidal mixture solution. This was because of higher index contrast and larger filling fraction of micro-particles in case of photonic glass. Longer fluorescence lifetime of embedded dye was observed in photonic glass environment in comparison to ethanolic solution of the Rhodamine-B dye. Lasing results obtained for photonic glass were compared with our previous results of Bloch lasing in photonic crystal synthesized using similar dye doped polystyrene micro-spheres.

1. Introduction:

In nano-photonic research a key priority has been to develop highly efficient nano-size threshold-less laser systems. Minimising size of a conventional optical cavity to nano scale and obtaining lasing with a tiny gain medium is a challenging task. A system with small gain requires extremely capable feedback mechanism for amplification and lasing. Progress towards development of micro-cavity lasers got a boost after invention of photonic crystal (PhC) in 1987 [1-2]. A typical PhC consisting of periodically arranged mono-dispersed spheres of diameter comparable to wavelength of incident radiation can control and manipulate propagation of light through it. An appropriate choice of lattice parameters and index contrast makes these structures a promising element to influence and provide feedback for emission from a fluorophore when embedded in this PhC environment [3-4]. In last two decades, several groups have reported lasing in colloidal PhCs implanted with several highly efficient luminescent components such as laser dyes [4-6], semiconductors [7], and rare earth ions [8]. In one of our earlier report, micro-cavity lasing was obtained in PhC synthesized using rhodamine-B (Rh-B) dye doped polystyrene (PS) spheres. Stimulated emission of the dye was obtained at 585 nm with a lasing threshold of 7.8 mJ pump energy in a well ordered highly reflecting (~70%) PhC [4]. Efforts toward developing such small volume or miniature lasers gained momentum in a new direction with the advent of random laser (RL).

Random lasing is a process of light amplification via multiple scattering owing to inherent randomness existing in a high- gain disordered medium. This phenomenon was theoretically predicted by Letokhov in 1968 [9] and subsequently demonstrated in various materials ranging from ceramic [10], semiconductor powder [11], nanostructure [12], dye solution [13], cholesteric crystal [14], quantum dot [15] and biological tissues [16].Two crucial conditions for RL are; availability of high gain and large level of scattering in the medium enabling emission to amplify and grow exponentially [17]. Unlike conventional feedback in lasers, multiple scattering of light provides feedback in a disordered medium. A diffused propagation of light in a random medium provides incoherent feedback and corresponding lasing results in the form of a smooth spectral distribution. However, when feedback occurs from interference of light, it is called coherent feedback and lasing is obtained in the form of several narrow emission lines or spikes [11]. After Garcia *el. al.* reported an innovative 3-dimensional system, called photonic glass (Ph-G) [18], designs for RL based miniature lasers moved towards Ph-G samples. Similar to PhC, Ph-G are also composed of mono-dispersed polymeric spheres of size comparable to the wavelength of light but unlike PhC these spheres are arranged in a completely disordered manner in Ph-G. Simple configuration and facile preparation make Ph-G useful for interesting applications such as, white paint [19], coherent back scattering [20], and random lasers. Garcia *et al* reported RL in an active ZnO Ph-G [21], Cerdan et al presented RL in dye doped latex Ph-G [22], Chen *et al* showed RL from π- conjugated polymers infiltrated in Ph-G [23], and Gottardo *et al.* reported RL in PS based Ph-G embedded with DCM dye [17].

Here we report random lasing from Rh-B dye attached to PS micro-particles. For our study, commercial aqueous colloidal suspensions of bare and Rh-B dye doped PS (PSRhB) spheres having diameter 295nm and 304nm, respectively, were used. Two random environments were generated using these spheres. In the first case, PS and PSRhB micro-particle suspensions were mixed to form binary colloidal mixture solution. In the second case, Ph-G samples were synthesized on glass substrates by drying these colloidal mixture solutions in ambient conditions. Mismatch in the micro-particle properties and their sphere diameters restricted periodic arrangement of spheres and disordered Ph-G were formed on the substrate. The PSRhB micro-spheres served as gain medium and the bare PS spheres performed as scattering centres in both random media reported in this study. A number of mixture solutions and Ph-G samples containing different volume fractions of PS and PSRhB spheres were examined. Since Rh-B is an efficient absorber at 532 nm, the samples were pumped using second harmonic of Nd:YAG laser and emissions were recorded at different pump energies using optical fibre based spectrometer. Gain narrowing appeared in both binary colloidal solution and Ph-G when the number density of PSRhB spheres was typically about $4.8 \times 10^8$ particles/µL in the sample. Lasing threshold in Ph-G was observed to be lower than the lasing threshold obtained in case of colloidal mixture solution. Our present results on Ph-G were also compared with the results obtained in our earlier studies for PhC synthesized by inward growing self-assembly technique using PSRhB spheres. In that case, the optical feedback occurred via reflections from Bragg planes of the crystal [4]. Fluorescence lifetimes of Rh-B dye in different environments such as, ethanolic solution, Ph-G and in PhC have been compared. Lifetime of Rh-B dye was found to be longer in case of Ph-G and PhC matrices, in comparison to ethanolic dye solution. Due to non-uniform distribution of density of states in PhC, a wavelength dependent fluorescence decay of Rh-B was observed. In comparison to PhC, a Ph-G based RL is easier to set up and does not require precise synthesis procedure. For achieving high reflectance in PhC a large number of defect free periodic layers are an absolute must. While laser output in PhC was directional was observed to have a broad amplified spontaneous emission background, in case of Ph-G, lasing occurred over a narrow spectral line width and had no directional dependence. Therefore, our results demonstrate Ph-G based random laser as an easy to fabricate miniature laser with superior spectral bandwidth in comparison to their Ph-C counterparts.

2. **Experiment:**

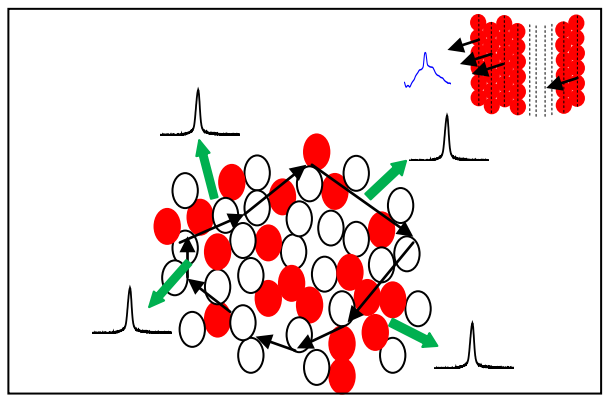

Figure 1: Schematic of RhB emission embedded in Ph-G, inset: in PhC

Aqueous colloidal solutions of bare and Rh-B dye doped PS spheres (dye concentration of 0.09 wt %) with particle diameters 295 nm and 304 nm, respectively, were commercially obtained from M/s Microparticles GmbH, Germany. Measured volumes of these two master solutions were mixed in pre-selected proportions to prepare sample mixture solutions for our study. Incorporating dye molecules into polymer beads has several advantages over a homogeneous dye solution. The polymer spheres not only provide protection to dye molecules against degradation and aggregation, but also make preparation of dye solution in any solvent (including solvents where dye solubility is low) possible without requiring additives. Additionally, possibility of self-quenching of dye emission resulting from insolubility and chemical reaction of dye is minimized in such an environment [24, 25]. For our study, standard 100 μL samples of PS and PSRhB colloidal suspensions used were estimated to have $5.48 \times 10^{11}$ and $1.69 \times 10^{11}$ spheres, respectively. RL characterizations were done in two types of disordered environment. In first case, a fixed volume of PS particles (300 μL) which corresponds to a total number of $1.6 \times 10^{12}$ spheres was mixed with PSRhB sphere suspension with volume ranging from 30 μL to 500 μL. Our previous studies had shown that number density of PS scatterers in the range of $2 \times 10^{12}$ particle/cm$^3$ provided optimum feedback for random lasing [16]. Hence, starting with a sample of bare PS sphere suspension having $5.48 \times 10^{12}$ particles/cm$^3$ systematic addition of dyed beads in the mixture was done to evaluate minimum number density of PSRhB particles which was required for lasing action to initiate and also to determine effect of number density of PSRhB particles on lasing. This binary colloidal mixture was housed in a quartz cuvette (optical path length of 1 cm) and sealed to avoid solvent evaporation. Lasing threshold and change in spectral width were measured as a function of the number density of the dyed micro-particles in the mixture.

Next, various Ph-G samples were synthesized using colloidal suspension mixture containing similar volume ratios of bare and dyed particles as studied in binary solutions in the first part of our investigations. A measured volume (250 μL) of each mixture was placed on a clean hydrophilic glass slide (1 cm x 1 cm) and left to dry in ambient. The solution when dropped on the glass slide formed a convex surface. Regions near the boundary of the drop were thinner than the centre and therefore dried faster. Resulting density gradient, eventually resulted in micro-particles flowing outwards and a large fraction of spheres settled at the periphery of the substrate. This resulted in a thick boundary of fabricated Ph-G. In preparation of Ph-G, the main aim was to prevent spheres from forming ordered layers. Difference in micro-particle properties and mismatch in their particle sizes prevented ordering and provoked colloidal flocculation of the spheres in the suspension and this resulted in cluster formation. After reaching a critical size, these clusters settled down arbitrarily and prohibited periodic self-assembly. Therefore, the resultant thin film generated contained a random distribution of disordered clusters of micro-particles. Multiple scattering from bare PS spheres provided optical feedback to the gain (Rh-B dye) present on PSRhB spheres and amplification occurred. A schematic of Ph-G containing bare (open circles) and dyed (filled circles) PS sphere with multiple scattering and multi-directional lasing is shown in Fig. 1. For comparison, inset in Fig. 1 shows a schematic of a typical PhC fabricated by inward-growing self-assembly technique using PSRhB spheres [4].

Both binary colloidal mixtures and Ph-G samples were pumped with second harmonic of Nd:YAG laser delivering pulses at 10 Hz (repetition rate), pulse duration of 6 ns and at wavelength of 532 nm. Emissions from the samples at $30^o$ with respect to normal were collected using an optical fibre based spectrometer.

Surface morphology of the Ph-G was examined under scanning electron microscope (SEM) and optical reflectivity of Ph-G sample was measured with a white light and an optical fibre based spectrometer.

For fluorescence decay lifetime measurements a time-correlated single-photon-counting spectrometer (1BH, UK) was used. In this case, the excitation source was a pulsed light emitting diode at 490 nm delivering 1.3 ns pulses at 1 MHz pulse repetition rate. Decay signals at various wavelengths were recorded using a photomultiplier tube. The observed fluorescence decay was analyzed by a convolution procedure using a proper instrument response function.

## 3. Results and discussion:

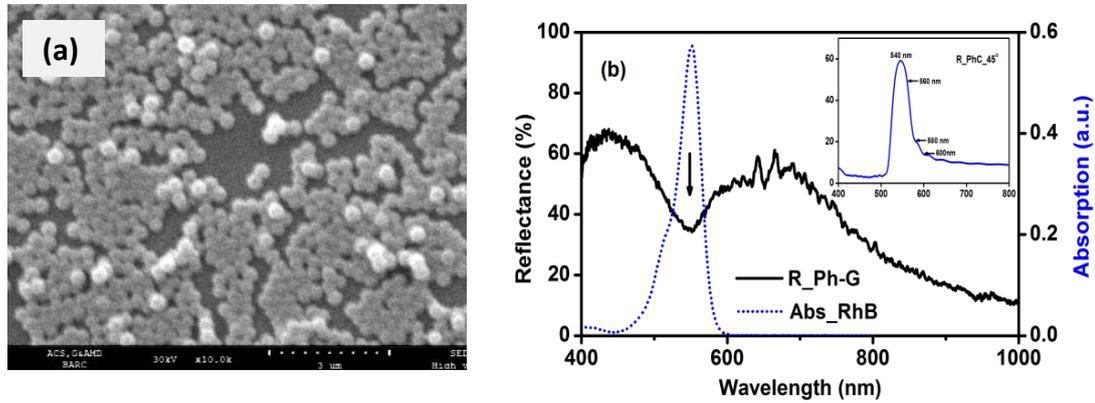

Figure 2: (a) SEM image showing top view of Ph-G, and (b) Reflection profile of Ph-G (solid line), absorption spectrum of Rh-B dye (dotted line), Inset: reflection spectrum of PhC (synthesized using PSRhB spheres) measured at 45°

Fig. 2a is a typical SEM image of top surface of a Ph-G sample. Arbitrary arrangement of micro-particles in the image confirmed random nature of the structure. Solid line in Fig.2b is the reflection spectrum of Ph-G recorded over a wavelength range of 400 nm to 1000 nm. The observed broad diffused profile of the reflection spectrum could be related to the random nature of the Ph-G microstructure. Such diffused propagating light through a randomly distributed gain volume creates possibility of incoherent feedback in the system [26]. The unusual dip in the reflection spectrum, shown with an arrow in Fig. 2b arises because of residual absorption by Rh-B dye molecules attached to some of the PS spheres. This is confirmed by the exact match between absorption spectrum of Rh-B shown by dotted line in the figure and the observed dip in reflection spectrum recorded for Ph-G.

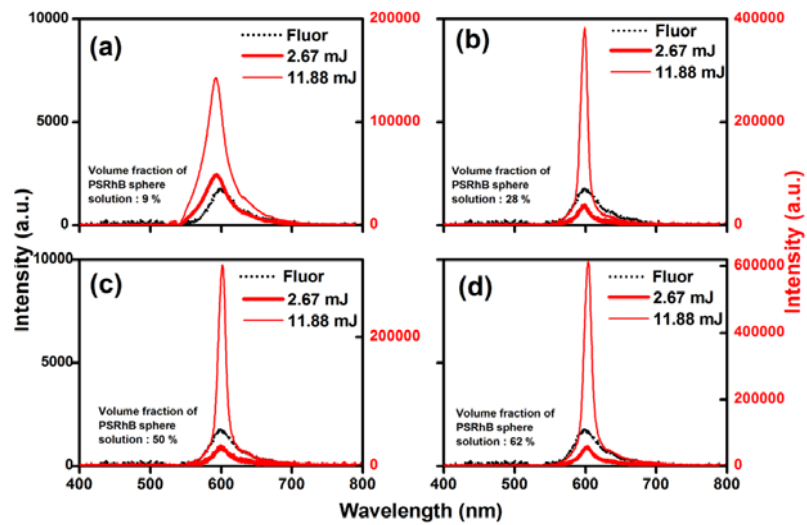

Figure 3: Fluorescence of PSRhB sphere suspension (dotted line), emission spectra of binary colloidal mixture solutions at 2.67 mJ (thick line) and at 11.88 mJ (thin line) containing PSRhB (PS) spheres with number densities: (a) $1.53 \times 10^7$ ($4.90 \times 10^9$) particles/µL, (b) $4.83 \times 10^8$ ($3.80 \times 10^9$) particles/µL, (c) $8.45 \times 10^8$ ($2.67 \times 10^9$) particles/µL, and (d) $1.05 \times 10^9$ ($2.05 \times 10^9$) particles/µL.

Binary colloidal mixtures containing bare and dyed micro-particles were pumped at different pump energies and emission spectra were collected at $30^o$ direction with respect to normal. Thick and thin lines in Figs. 3a-3d show the emission spectra of the mixture solutions at two extreme pump energies, 2.67 mJ and 11.88 mJ, respectively. For comparison, spontaneous emission of PSRhB sphere suspension is shown with dotted line in Fig. 3a-3d. Initially, 30 µL of PSRhB sphere solution was added to 300 µL of PS sphere suspension. This binary colloidal mixture solution containing ~ 9% of dyed spheres had $4.90 \times 10^9$ particles/ µL and $1.53 \times 10^7$ particles/ µL of PS and PSRhB spheres, respectively. The gain was not sufficient in this mixture and therefore, spectral narrowing was not observed even at high pump energy of 11.88 mJ, as shown in Fig. 3a. Fig. 3b shows the emission spectra of binary colloidal mixture containing ~ 28% volume fraction of PSRhB sphere solution and this corresponds to $4.83 \times 10^8$ particles/µL. For this, 120 µL of PSRhB spheres solution was added to 300 µL of PS spheres suspension; this resulted in an effective reduction in number density of scatterers to $3.80 \times 10^9$ particles/µL. The broad spontaneous emission of the dye observed at pump energy of 2.67 mJ got spectrally compressed into a sharp narrow emission for pump energy of 11.88 mJ in this case, as shown in Fig. 3b. The lasing peak was centered at 598 nm. Fig. 3c and 3d show emission spectra of binary colloidal mixture solutions containing 50% and 62% of volume fractions of PSRhB sphere solution, respectively. Addition of PSRhB sphere solution increased the number density of dye particles to $8.45 \times 10^8$ particles/µL (50%) and $1.05 \times 10^9$ particles/µL (62%). However, addition of PSRhB sphere solution automatically reduced the number density of PS sphere to $2.67 \times 10^9$ particles/µL (Fig. 3c) and $2.05 \times 10^9$ particles/µL (Fig. 3d) in the binary colloidal mixture. A red shift of lasing peak by 10 nm observed between Fig. 3b and 3d occurred largely due to self absorption because of the relatively larger number density of dyed particles in the solution.

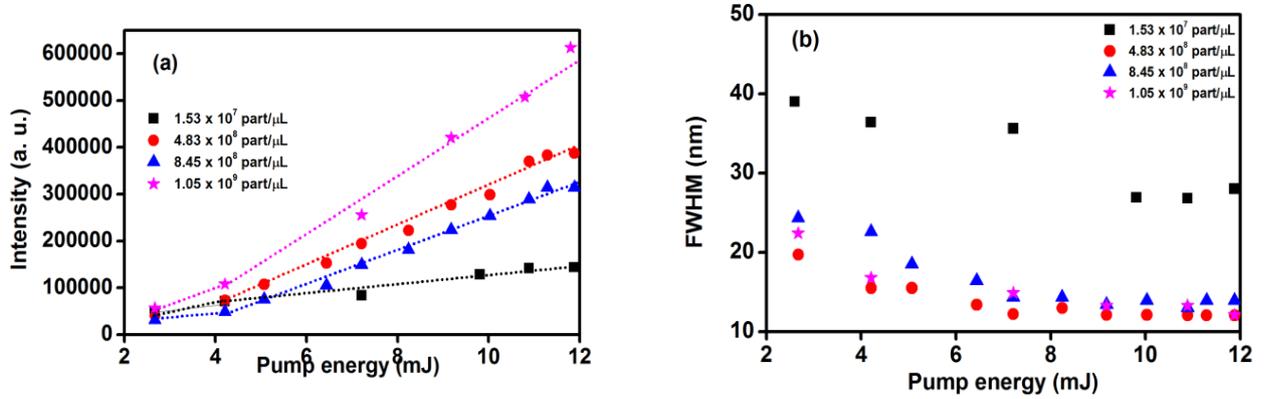

Figure 4: (a) Dependence of emission peak intensities, and (b) FWHM, on pump energy for binary colloidal mixture solutions containing different volume fractions of PSRhB sphere solution

Fig. 4a shows variation in the peak emission intensity from different binary colloidal mixtures with varying pump energies. All three suspension mixtures containing ≥ 28% volume fraction of PSRhB spheres showed lasing threshold of ~ 4.2 mJ, as shown with circles (28% of PSRhB sphere solution), triangles (50% of PSRhB sphere solution) and stars (62% of PSRhB sphere solution) in Fig. 4a. Larger number density of PSRhB spheres in the mixture solution did not improve lasing in terms of reduction in threshold. Addition of PSRhB spheres increased total volume of the suspension mixture in the cuvette and this resulted in simultaneous reduction in number density of scatterers. Addition of measured volumes of aqueous master solution containing PSRhB spheres does increase the number density of dye spheres in the suspension, resulting in enhanced gain which is expected to aid laser action in the medium. However, there occurs a simultaneous reduction in scatterer concentration from 90% to 37%, which has a counteractive effect through reduced feedback. The lasing threshold remained largely unchanged as shown in Fig. 4a. Effects arising from increased concentration of gain centers along with reduced concentration of scattering centers countered each other leading to a nearly similar spectral trend and spectral widths being observed in all these cases (Fig. 3b-3d). Lasing action in a colloidal solution consisting of only PSRhB spheres could not be observed in our present study, suggesting that feedback in such a medium was inadequate for random lasing to occur. Martin et al. did observe lasing in Rhodamine-6G doped PS sphere suspension but feedback in their case was provided by a 90% reflective aluminium mirror and the end lateral face of the quartz cell containing the suspension.

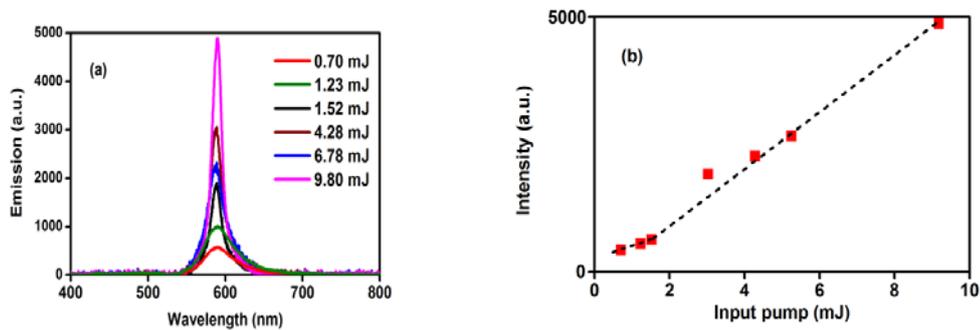

**Figure 5:** (a) Emission spectra of Ph-G samples (containing 50% volume fraction of PSRhB sphere solution) at different pump energies, and (b) Peak emission intensities as a function of pump energies.

In case of Ph-G, emission measurements were performed using the uniform thick peripheral region of the Ph-G samples described in Section-2. Fig. 5a shows, pump energy dependent emission spectra of Ph-G synthesized using colloidal mixture solution containing 50% of PSRhB colloidal solution. Spectral narrowing was observed for pump energy ~ 1.52 mJ and above. Gain narrowing with a smooth spectral profile consisting of single lasing peak indicated incoherent feedback in the system. Fig. 5b shows the excitation dependent enhancement in emission intensity of the sample. This sample of Ph-G showed a lasing threshold of ~1.5 mJ, which corresponds to a pump intensity of 1.91 W/cm$^2$, as shown in Fig. 5b. Cerdan *et. al.* reported RL in Ph-G containing only Rhodamine-6G doped polymeric nano-particles (without addition of any scatterers) at much higher lasing threshold of hundred of kWcm$^{-2}$ [22]. Hence, our results suggest that addition of scatterers in the Ph-G disordered medium facilitates efficient random lasing by providing necessary feedback.

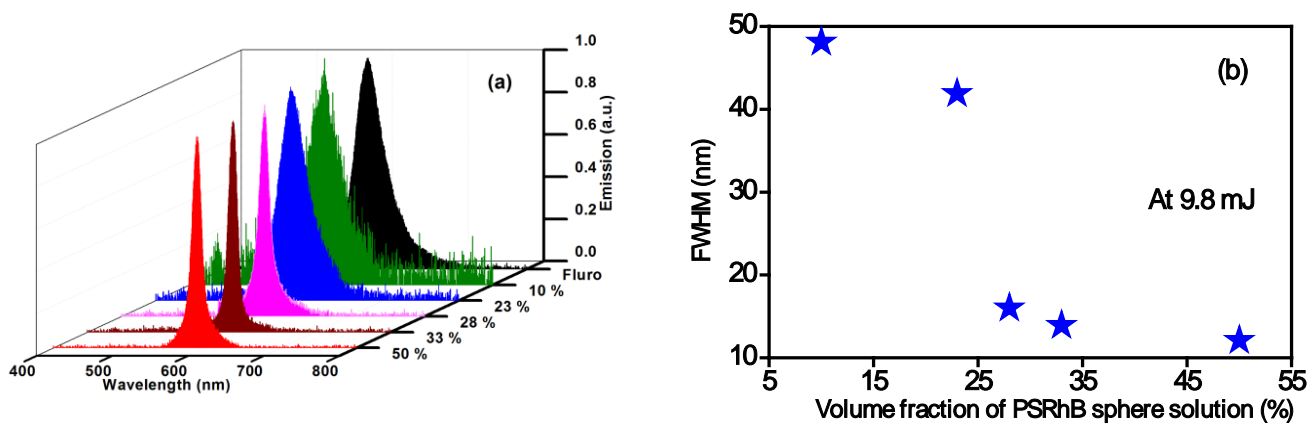

**Figure 6:** (a) Emission recorded at 9.80 mJ pump energy for Ph-G samples containing different volume fraction of PSRhB micro-particles, and (b) FWHM of emission spectra with respect to increased volume fraction of PSRhB sphere solution

Emissions recorded at pump pulse energy of 9.80 mJ from a number of Ph-G samples prepared with colloidal mixtures containing different volume fraction of aqueous suspension of PSRhB spheres are shown in

Fig 6a. For small volume fractions such as 10% and 23% of PSRhB sphere suspension, the emission of Ph-G did not differ much in comparison to fluorescence of the dye (depicted in black). However, a notable narrowing in the spectral width was observed for Ph-G made with the binary solution containing volume fraction of 28% ($4.83 \times 10^8$ number/µL) and above of PSRhB particle suspension. Similar volume ratios of bare and dyed particles in binary colloidal mixture solution too had shown spectral narrowing (Fig. 3b). This suggests, on an average, about 10 times more number of scatterers were required in comparison to dyed beads to obtain spectral narrowing. In comparison to binary colloidal solution, a marginal reduction in spectral width was observed in Ph-G on addition of (33 % & 50 %) PSRhB colloidal solution. Fig. 6b shows FWHM of emission spectra recorded at 9.80 mJ pump energy for Ph-G samples containing different volumes of PSRhB spheres.

Lasing threshold in Ph-G (1.5 mJ) was observed to be lower than threshold in binary colloidal solution (4.2 mJ). The main difference between binary suspension and Ph-G was the environment seen by the dye molecules in these two cases. The lower lasing threshold in Ph-G could arise most probably due to combined effect of larger filling fraction of the dyed beads, hence higher gain per unit length, and a high refractive index contrast between PS and air which provided stronger scattering effects. The index contrast in case of binary suspension solution between water and PS being much less than index contrast between PS and air effective in Ph-G, stronger scattering is expected in Ph-G.

Although, any arbitrary distribution of poly-dispersed particles, grains or clusters can form a random environment, in Ph-G monodispersity of particles leads to a resonant behaviour. A suitable choice of diameter and refractive index of the mono-dispersed spheres can result in scattering resonance within the gain wavelength range, thereby, providing a control over the lasing wavelength [27]. Additionally, mono-dispersity provides uniform filling fraction in the sample ensuring reproducible scattering length and emission property across the sample [28]. Ph-G is distinguished from PhC by their significantly different structural and optical properties. Mechanism of optical feedback by scattering in Ph-G is much different from the mechanism of feedback in case of PhC [4]. Micro-cavity lasing from PhC synthesized using PSRhB micro-particles with sphere diameter 302 nm has been reported [4]. These PhCs possessed a clear photonic stop band in ΓL direction. The de-excited photons which spectrally matched the photonic stop band faced reflections from Bragg planes of the crystal and were efficiently trapped and localized within the PhC. This resulted in increased effective feedback and residence time of the photons within the PhC matrix. These photons which remained trapped in a particular direction emerged from other directions with enhanced emission intensity. A schematic of PhC matrix with multiple Bragg reflections is shown in the inset of Fig. 1. For a PSRhB microspheres based PhC with reflectance

~70 %, stimulated emission of the dye dominated and lasing was observed at 22º with respect to the normal. Lasing peak at 587 nm with FWHM of 9 nm and a lasing threshold of 7.3 mJ was reported [4]. The lasing wavelength exactly matched with the photonic band edge and the stimulated emission observed in case of PhC was accompanied by a broad amplified spontaneous emission band. Larger density of states available at the photonic band edge supported Bloch lasing in PhC. The lasing cavity was further improved by gold coating the top surface of the PhC [29].

As reported earlier, the fluorescence decay lifetime of RhB dye was observed to be longer in PhC environment in comparison to its lifetime in ethanolic solution [30]. In PSRhB micro-particle based PhC when recorded at 590 nm, Rh-B dye showed bi-exponential decay with a longer decay component of 4.09 ns. This value was larger than decay lifetime of the dye in solution ~ 2.73 ns [30]. Due to larger density of states available at PhC band edge, a faster decay of Rh-B for wavelengths matching the photonic band edges was expected. In present study, we compared the fluorescence decay lifetime obtained at different wavelengths of Rh-B dye when embedded in: (a) Ph-G and (b) PhC matrices.

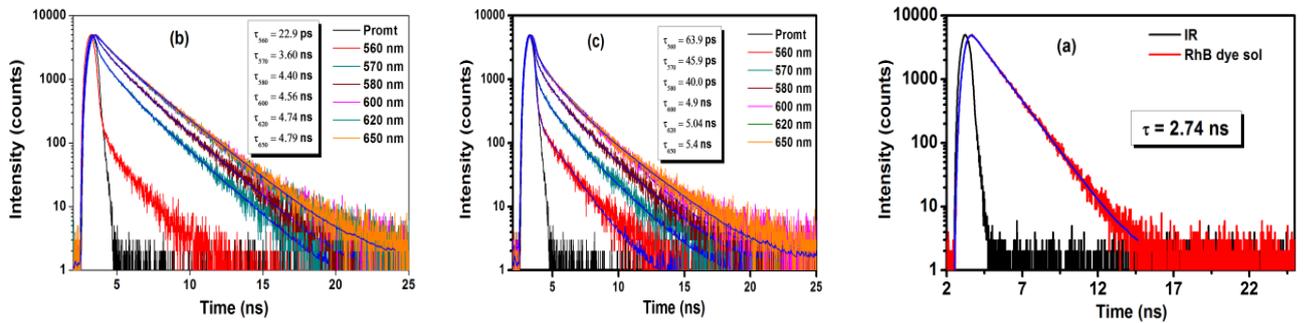

**Figure 7:** Excited state life time of RhB (a) in ethanol, (b) in Ph-G recorded at different wavelengths, and (c) in PhC recorded at different wavelengths

Fig. 7a depicts our results on fluorescence decay of Rh-B dye dissolved in ethanol. The fluorescence of the dye exhibited exponential decay with fluorescence decay lifetime of 2.74 ns. In Fig. 7b and 7c are shown fluorescence decay of RhB dye in Ph-G and PhC samples, respectively, measured at different wavelengths. In both the cases, the decay traces could be fitted by multi-exponential components. The major decay component measured at different wavelengths in the case of Ph-G and PhC are summarized in Table-I. Decay lifetimes of the dye in Ph-G were found to be nearly similar at all measured wavelengths. However, in case of PhC, the lifetimes at wavelengths 570 ns and 580 ns which lie close to the photonic band edge were found to lie in the range of tens of picoseconds. For wavelengths far from stop band of the Ph-C sample such as at 600 nm, 620 nm and 650 nm, measured lifetime was in the range of a few nanoseconds as expected. For reference, the measured photonic stop band for PSRhB based PhC sample under study, measured at 45° with respect to normal, is shown in the inset of Fig. 2b.

Table-I: Wavelength dependent fluorescence lifetime of Ph-G and PhC

| Wavelength | 570 nm | 580 nm | 600 nm | 620 nm | 650 nm |
| --- | --- | --- | --- | --- | --- |
| Decay in Ph-G | 3.60 ns | 4.40 ns | 4.56 ns | 4.74 ns | 4.79 ns |
| Decay in PhC | 45.9 ps | 40.0 ps | 4.9 ns | 5.04 ns | 5.40 ns |

A considerable variation in fluorescence lifetime of Rh-B dye attached to PS sphere was observed in comparison to Rh-B dye solution in ethanol. The rigid architecture of PS provided a stiff environment to Rh-B dye molecules therefore restricting molecular mobility hence reducing the probability of non-radiative decay pathways [24]. Multi-exponential decay nature of Rh-B in PhC and Ph-G environment can be attributed to sub-ensembles of dye molecules undergoing different interaction within their solid environment [31]. Since the fluorescing dye in both PhC and Ph-G possess similar PS environment, shorter decay at the photonic band edge and variation in lifetime with wavelength in case of PhC confirmed the effect of photonic stop band or the available density of states on the fluorescence decay lifetime of the embedded dye.

We have observed a lower lasing threshold in Ph-G in comparison to PhC. Our results demonstrate Ph-G to be a competitive medium for miniature lasers. Easy to fabricate, such Ph-G samples offer, high diffuse reflection over a broad wavelength range in presence of suitable embedded scatterers enabling emission over a broad wavelength range to be achieved. In contrast, precise synthesis with a demand for large number of defect free periodic layers limit use of PhC. Lasing action in PhC is determined mainly by its stop band wavelength and band gap effects hence restricting the emission spectrum. Hence, RL in a disordered medium is an easier and more effective route to achieving miniature lasers than Bloch lasing in periodic micro-cavity. On the other hand,

4. **Conclusion:**

Random lasing was obtained in binary colloidal solution and in Ph-G containing a mixture of bare and dyed PS micro-particles in different volume ratios. Randomly positioned bare PS spheres effectively served as scattering centres and in the absence of a conventional optical cavity both disordered media worked as small volume lasers. Because of higher refractive index contrast and larger filling fraction in Ph-G, the lasing threshold in Ph-G was lower in comparison to binary colloidal mixture solution. In contrast to directional lasing associated with broad amplified spontaneous emission background observed in PhC, lasing with narrow spectral line width was obtained in Ph-G in all directions over a wide wavelength range. Thus, Ph-G based random laser offers a potential alternative for achieving micro-cavity lasers in comparison to Bloch lasing in PhC.

**References:**

[1] E. Yablonovitch, "Inhibited spontaneous emission in solid-state physics and electronics," *Phys. Rev. Lett.,* 58 (1987) 2059

[2] S. John, "Strong localization of photons in certain disordered dielectric superlattices," *Phys. Rev. Lett.,* 58 (1987) 2486

[3] Y. Yamamoto and R. E. Slusher, *Phys. Today* **46** (1993) 66.

[4] S. Kedia, R.Vijaya, A.K.Ray and S. Sinha, *J. Nanophotonics* **4** (2010) 049506.

[5] N. Eradat, M. Wohlgenannt, Z. V. Vardeny, A. A. Zakhidov, and R. H. Baughman, Synth. Met. 116 (2001) 509.

[6] K. Yoshino, S. B. Lee, S. Tatsuhara, Y. Kawagishi, M. Ozaki and A. A. Zakhidov, Appl. Phys. Lett. 73 (1998) 3506.

[7] A. K. Khokhar, R. M. De La Rue, and N. P. Johnson, IET Circuits Devices Syst. 1 (2007) 210.

[8] M. Li, P. Zhang, J. Li, J. Zhou, A. Sinitskii, V. Abramova, S. O. Klimonsky, and Y. D. Tretyakov, Appl. Phys. B. 89 (2007) 251

[9] V. S. Letokhov, Sov. Phys. JETP 26, 835 (1968)

[10] M. Bahoura, K. J. Morris and M. A. Noginov, Opt. Commun. 201, (2002) 405-411.

[11] Dominguez CT, Gomes MDA, Macedo ZS, Araujo CBD, Gomes ASL. Multiphoton excited coherent random laser emission in ZnO powders. Nanoscale 7 (2015) 317

[12] Kumar B, Patel SKS, Gajbhiye NS, Thareja RK. Random laser action with nanostructures in a dye solution. J Laser Appl 25 (2013) 042012-1

[13] S. Kedia and S. Sinha, Results in Physics 7 (2017) 697

[14] Li L, Deng L. Low threshold and coherent random lasing from dye-doped cholesteric liquid crystals using oriented cells. Laser Phys 23 (2013) 085001-1.

[15] Chen Y, Herrnsdorf J, Guilhabert B, Zhang Y, Watson IM, Gu E, Laurand N, Dawson MD. Colloidal quantum dots random laser. Opt Express 19 (2011) 2996

[16] R. C. Polson and Z. V. Vardeny Appl. Phys. Lett. 85 (2004) 1289

[17] S. Gottardo, R. Sapienza, P. D. Garcia, A. Blanco, D. S. Wiersma and C. Lopez, Nature Photonics 2 (2008) 429

[18] P. D. Garcia, R. Sapienza A. Blanco and C. Lopez, *Adv. Mater.* 19 (2007) 2597.

[19] P. D. Garcia, R. Sapienza and C. Lopez, *Adv. Mater.* **22** (2010) 12.

[20] Y. Kuga, A. Ishimaru, J. Opt. Soc. Am. A 1984, 8, 831

[21] P. D. Garcia and C. Lopez, J. Mater. Chem. C. 1 (2013), 7357

[22] L. Cerdan, A. Costela, E. Enciso, and I. Garcia-Moreno, Adv. Funct. Mater. 23(2013), 3916

[23] Chen Y, Herrnsdorf J, Guilhabert B, Zhang Y, A. L. Kanibolotsky, P. J. Skabara, Gu E, Laurand N, Dawson MD Organic Electronics 13 (2012) 1129

[24] V. Martin, J. Banuelos, E. Enciso, I. L. Arbeloa, A. Costela, and I. Garcia-Moreno, J. Phys. Chem. C. 115 (2011) 3926

[25] E. Enciso, A. Costela, I. G. Moreno, V. Martin, and R. Sastre, Langmuir, 26 (2010) 6154

[26] H. Cao, Waves Random Media, 13 (2003) R1

[27] M. I. Antonoyiannakis, R. Pendry *Europhys. Lett.* 1997, 40, 613

[28] Chen Y, Herrnsdorf J, Guilhabert B, Zhang Y, A. L. Kanibolotsky, P. J. Skabara, Gu E, Laurand N, Dawson MD Organic Electronics 13 (2012) 1129-1135

[29] S. Kedia, R.Vijaya, A.K.Ray and S. Sinha, *Opt. Commun.* 284 (2011) 2056.

[30] S. Kedia and S. Sinha, *J. Phys. Chem. C.* 119 (2015) 8924

[31] Barth, M.; Gruber, A.; Cichos, F. Spectral and Angular Redistribution of Photoluminescence Near a Photonic Stop Band. Phys. Rev. B 2005, 72, 085129−1−10.